\documentclass[preprint]{aastex}
\usepackage{graphicx}

\begin{document}

\title{On the Possibility of Identification of a Short/Hard Burst GRB~051103
with the Giant Flare from a Soft Gamma Repeater in the M81 Group of
Galaxies }

\author{D. D. Frederiks\altaffilmark{1}, V. D. Pal'shin\altaffilmark{1},
 R. L. Aptekar'\altaffilmark{1}, S. V. Golenetskii\altaffilmark{1},
 T. L. Cline\altaffilmark{2}, and E.~P.~Mazets\altaffilmark{1}}
\altaffiltext{1}{Ioffe Physical-Technical Institute, St.Petersburg,
194021, Russia.}
\altaffiltext{2}{Goddard Space Flight Center, NASA,
Greenbelt, MD 20771, USA.}


\begin{abstract}
The light curve, energy characteristics, and localization of a
short/hard GRB~051103 burst are considered. Evidence in favor of
identifying this event with a giant flare from a soft gamma repeater
in the nearby M81 group of interacting galaxies is discussed.
\end{abstract}

\section{Introduction}

Observations of giant flares from SGR~1900+14 on August 27, 1998,
and from SGR~1806-20 on December 27, 2004 renewed discussions of an
old suggestion that some of short/hard gamma-ray bursts may
represent initial pulses of giant flares from considerably more
remote gamma repeaters \citep{Mazets81,Mazets82}. The energy of the
initial pulse of the giant flare from SGR~1806-20 was found of
$\sim(2$--$3)\times 10^{46}$~erg, making it possible to estimate the
maximum distance at which flares of this kind can still be revealed
as short bursts to be 30--50 Mpc \citep{Mazets05, Hurley05,
Palmer05, Terasawa05, Mereghetti05}. Such estimates give reason to
believe that giant flares can be detected not only in galaxies of
the Local group, but also in nearby galaxies and clusters of
galaxies, such as, e.g., clusters in Virgo, Ursa Major, Leo.

The available data on giant flares in three SGRs originated in the
Milky Way outline the expected temporal and spectral characteristics
of short gamma-ray bursts that could be indicative of a flare from
another galaxy. Such a burst must have the shape of a single pulse
with a steep leading edge ($\sim$~5--15~ms) and quasi-exponential
decay with $\tau \sim$50--70 ms at a total pulse width of
$\sim$200--300~ms. Of course, such an idealized light curve will
actually be distorted by noticeable faster fluctuations of the
emission intensity. The energy spectrum of a burst must be rather
hard in the initial part of the pulse and extend to 10--15 MeV. In
the course of the burst, the spectrum rapidly evolves to become
considerably softer by the end of the pulse. Although, at first
sight, these specific features do not make cardinal differences from
a wide variety of short/hard gamma-ray bursts observed, a closer
inspection of the Konus catalog of short bursts accumulated in
1994--2002 \citep{Mazets02}, suggests that the fraction of events
with the characteristics specified does not exceed several percent
of the total number of short bursts. Naturally, a decisive argument
in favor of identification of the short burst with a giant remote
flare would be that a galaxy with a redshift of $z \leq 0.01$ is
found in the region of its localization.

It is widely believed that neutron stars passing through the SGR
stage are young. Their age estimated e.g. by the secular variation
of the period, $T \sim \frac{1}{2} P/\dot{P}$, does not exceed $\sim
10^4$~yr. In accordance with the assumption that the most probable
candidates in a search for SGR are galaxies with a high rate of
formation of massive stars, Popov and Stern (2006) subjected to a
statistical analysis the BATSE data on short gamma-ray bursts. They
concluded that the fraction of the possible extragalactic flares
among the short bursts recorded by BATSE does not exceed several
percent and the best target to search for giant flares is a cluster
of galaxies in Virgo. \citet{Nakar06} estimated the fraction of SGR
flares relative to all short/hard bursts to be $< 15\%$.

On November 3, 2005 the Konus-Wind gamma-burst spectrometer detected
an intensive short/hard gamma-ray burst. The GRB~051103 was also
observed by the spacecraft HETE-2 (FREGATE instrument), Mars Odyssey
(GRS and HEND), Swift (BAT, outside the field of view), and RHESSI
(PCA). As a result, this interplanetary network (IPN) localized the
burst source. The center of the error box has coordinates $\alpha =
9^{\mathrm h} 52^{\mathrm m} 32^{\mathrm s}$, $\delta = 68\arcdeg
50\arcmin 42\arcsec$ (J2000) \citep{Golenetskii05}. The area of the
error box (at the $3 \sigma$ level) is 260 sq.~arcmin. The
localization area lies close to the M81 galaxy but does not overlap
with its optical image. The box center is 21~arcmin away from the
center of M81.

\citet{Lipunov05} reported that the localization region definitely
lies outside the spiral arms of the M81 galaxy, but noted that the
structure of the galaxy is appreciably distorted by the tidal
interaction. If GRB~051103 is not related to the flare in SGR, it
would be expected, with a certain probability, that an optical
transient (OT) should appear in the IPN box. A search for an OT has
been performed, but yielded only the upper limit to its brightness:
$m > 18.5$ \citep{Lipunov05}, $m_R > 21$ \citep{Klose05}, and $m_R
> 19.5$ \citep{Ofek05}. A search for high-energy emission from
the IPN box during a burst of length $\Delta T = 0.17$~s was
performed on the Milagro gamma-ray observatory. Also, only the upper
limit to the integral fluence was obtained: $< 4.2 \times
10^{-6}$~erg~cm$^{-2}$ ($\Delta E = $0.25--25 TeV)
\citep{Parkinson05}. If only these results are considered,
identification of GRB~051103 with the giant flare from an SGR in M81
seems to be hardly probable.

The group of close interacting galaxies M81 (D=3.6~Mpc), which
includes several galaxies and about ten smaller stellar structures,
has also been studied during many years by means of radio and X-ray
astronomy. Results of these studies taken into account change the
situation. There appear important arguments indicating that the
probability of association of GRB~051103 with the M81 group is
rather high. This evidence will be considered further in discussion.

\section{Observations and localization}
\begin{figure}
\centering
\includegraphics[]{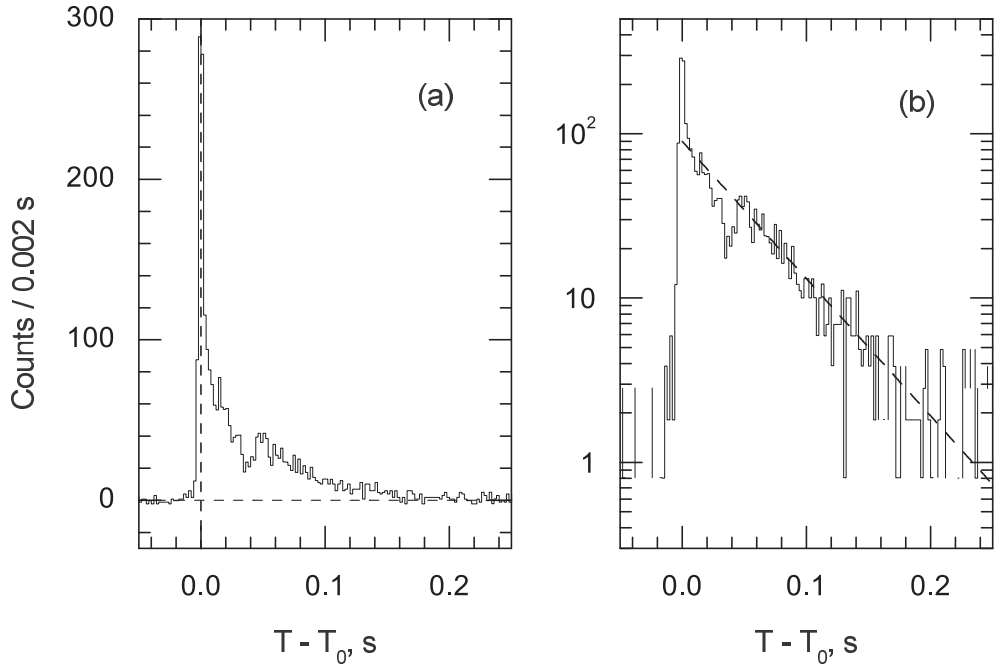}
\caption{Light curve of GRB~051103 in a wide energy range 18--1100
keV (a). The burst decays quasi-exponentially. Dashed line
corresponds to $\tau = 55$~ms (b). \label{th1}}
\end{figure}

\begin{figure}
\centering
\includegraphics[]{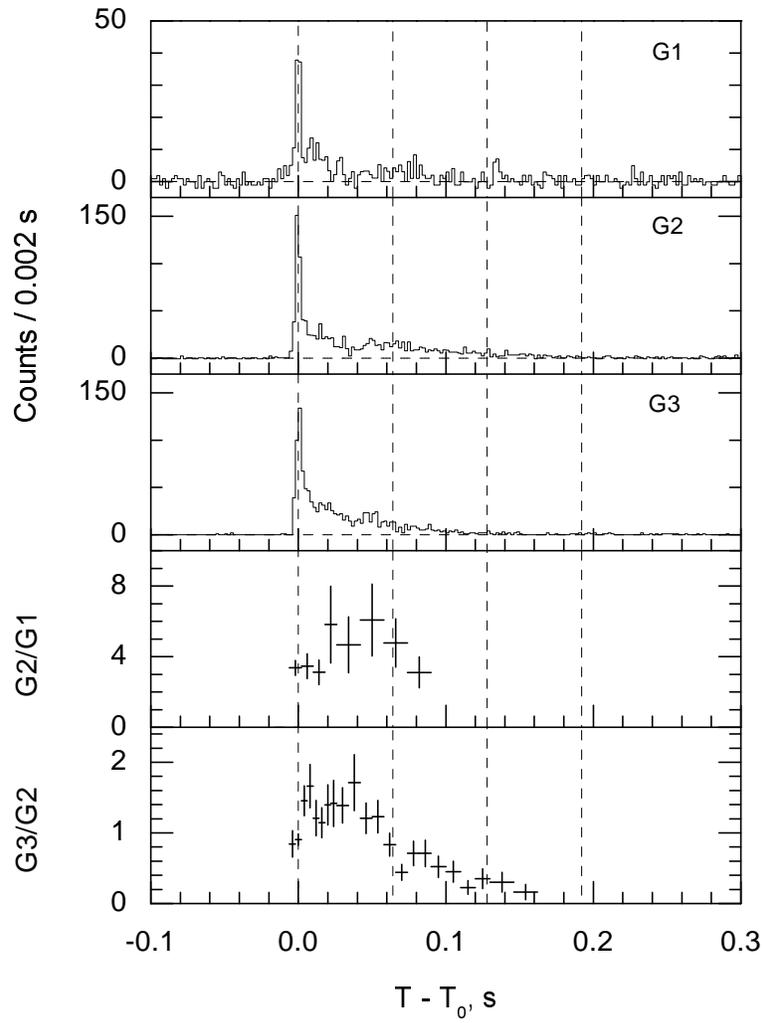}
\caption{Time history of the burst, recorded in three energy ranges:
G1(18--70 keV), G2(70--300 keV), G3 (300-1100 keV) and the hardness
ratios. Vertical dashed lines denote the time intervals of energy
spectra accumulation. \label{th2}}
\end{figure}

The light curve of GRB~051103, recorded in a wide range of energies
(18--1100 keV) with a resolution of 2 ms, is shown in Fig.~\ref{th1}
in linear (a) and semilog (b) scales. The burst has the form of a
single pulse with a steep leading edge ($\leq 6$ ms) and a
quasi-exponential decay ($\tau \sim 55$~ms). The total duration of
the burst is $\sim 170$~ms. The time profiles for three energy
windows, G1 (18--70 keV), G2 (70--300 keV), and G3 (300--1100 keV),
and the hardness ratios G2/G1 and G3/G2 are shown in Fig.~\ref{th2}.

The variations of the hardness with time and especially the run of
G3/G2 indicate that the spectrum shows a clearly pronounced
evolution and rapidly becomes softer. In the course of a burst,
three multichannel spectra were measured with an accumulation time
of 64 ms for each spectrum. To improve a poor statistics of counts
the weaker spectra 2 and 3 were summarized. The response matrix of
the detector was calculated for the known incident angle of
radiation. The photon spectra of the burst were obtained by fitting
their intensity and shape to the instrumental energy loss spectra
using XSPEC v. 11.3.

Spectrum 1 covers the most intense and hardest part of the burst.
Emission is seen up to 10 MeV. The spectrum is well fit by a
power-law distribution with an exponential cutoff $dN \propto
E^\alpha \exp(-E/E_0)dE$, which transforms at high energies into a
decaying power-law tail $dE \propto E^{-\beta}dE$ (Band model). The
parameters have the following values:
$$
\alpha = 0.16^{+0.19}_{-0.15} ; \hspace{6mm} E_0 =
1050^{+270}_{-200} ; \hspace{6mm} \beta = 2.9^{+0.5}_{-1.7} ;
\hspace{6mm} (\chi^2 = 38/35\,dof).
$$

Here and henceforth the errors in the parameters and energy
estimates correspond to a confidence probability of 90\%. It should
be noted that, in contrast to the most of short/hard bursts, the
spectra under consideration have $\alpha > 0$. The peak energy of
the $\nu F_\nu$ spectrum, $E_p = 2300_{-150}^{+350}$~keV. The energy
fluence for the time interval $\Delta T = 64$~ms is $S = (3.4 \pm
0.4) \times 10^{-5}$~erg~cm$^{-2}$. The 2-ms peak flux, $F_{max} =
(2.8 \pm 0.3) \times 10^{-3}$~erg~cm$^{-2}$~s$^{-1}$.

\begin{figure}
\centering
\includegraphics[]{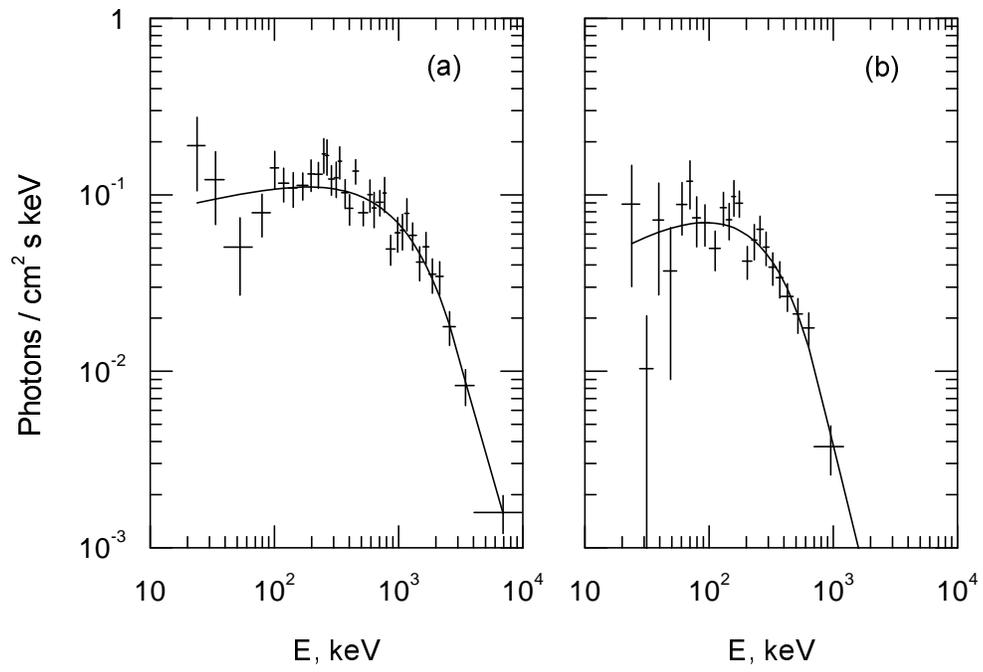}
\caption{Photon spectra of the burst, measured in the time intervals
0--64~ms (a) and 65--192~ms (b). \label{sp}}
\end{figure}

The sum spectrum (2+3) characterizes the final stage of the burst.
Emission is observed up to $E \sim 2$~MeV. The spectrum is well
described by a power-law model with an exponential cutoff
(Fig.~\ref{sp}b) with $\alpha = 0.43^{+0.34}_{-0.40}$, $E_0 =
220^{+70}_{-50}$~кэВ ($\chi^2 = 21/20\,dof$), $E_p = 530 \pm
80$~keV. The fluence in the interval $T - T_0 =$65--192 ms is $S =
(2 \pm 0.3) \times 10^{-6}$~erg~cm$^{-2}$.

With account taken of the radiation arrived before $T_0$
(Fig.~\ref{th1}) and on the assumption that the spectrum of this
radiation is close to spectrum 1, the total fluence of the whole
burst is $S = (4.4 \pm 0.5) \times 10^{-5}$~erg~cm$^{-2}$.

For triangulation of the source of GRB~051103 the positions of the
most remote spacecraft Mars Odyssey and Wind spacecraft (distance
from the Earth 1.6 million kilometers), and the position of
near-Earth orbiting Swift spacecraft were chosen as the main
vertices of IPN. The choice among the two alternative localization
regions, represented by intersections of the triangulation rings of
Mars Odyssey and Wind--Swift was based on the fact that the S2
detector on Wind, which recorded the burst, scans the northern
hemisphere over the ecliptic. Data of other near-Earth spacecraft,
HETE-2 and RHESSI confirm this localization.

The coordinates of the center and corners of the $3 \sigma$ error
box are listed in Table 1.

\section{Discussion}
The near group of galaxies M81 is constituted by six principal
members with luminosities $L/L_\odot$ of $2 \times 10^{10}$ to
$10^8$ \citep{Yun99}. The central part of the group is represented
by three galaxies: M81, M82, and NGC~3077. The M81 group has been
intensively studied during many years. While observations in the
visible band gave only little sign of tidal interactions in this
group, studies by VLA at a wavelength of 21~cm yielded convincing
evidence of the strong influence of tidal forces on the structure of
the group \citep{Yun99}. \citet{Yun94} discovered that the three
central components of the M81 group are embedded in a cloud of
atomic hydrogen with a mass of $\sim 5 \times 10^{9}M_{\odot}$.
About one fourth of this mass is distributed in the space between
the galaxies in the form of filaments, spirals, and bridges
connecting the galaxies. The complex structure of the distribution
of densities and radial velocities of HI was formed upon a close
approach of galaxies, accompanied by intensive disintegration of
their outer regions by the tidal interaction. \citet{Yun99}
performed a numerical modeling of the galaxy interaction within the
capabilities of the restricted three-body problem. According to this
simulation, the galaxies collided about 300 million years ago. The
disks and gas envelopes of the less massive galaxies NGC~3071 and
M82 suffered the strongest disruptions $2.8\times10^8$ and
$2.2\times10^8$ years ago correspondingly. These estimates are in an
agreement with the other available estimates of the time when star
formation processes were enhanced in disintegrated regions.

Irregular dwarf galaxies, most probably of tidal origin, have been
observed in the maximum-density HI regions. \citet{Makarova02} using
the Hubble Space Telescope obtained color--magnitude diagrams for
four dwarf galaxies Holm IX, BKN3, Arp-loop, and Garland. Main
sequence stars, bright blue stars, red supergiants, and red giants
are represented in these diagrams in various ratios. A study of the
diagrams demonstrated that the process of star formation at a rate
of $\sim 7.5 \times 10^{-3}$~--~$7.5 \times10^{-4}
M_{\odot}$~yr$^{-1}$ occurred in the time interval from 20 to 200
million years ago. It is natural to assume that stars were also
formed, even though at a lower rate, in other, less dense HI
regions.

\citet{Sun05} performed a deep spectral survey of the integral
stellar emission in the region M81/M82/NGC~3077. They discovered a
widely distributed stellar population in the space between the
galaxies. The distribution of the emission intensity follows the
structure of atomic hydrogen clouds. The emission spectra of this
stellar population are the same to the east and west of M81 and to
the south of M82. This may mean that either the stars had belonged
to the disk of M82 destroyed upon close approach to M81 several
hundred million years ago or were formed in dense HI regions.

The Chandra X-ray observatory performed a survey of the M81 galaxy
and its close vicinity \citep{Swartz03}. As a result, 177 X-ray
sources were discovered, which mainly are accreting binary systems
with a luminosity $\gtrsim 10^{37}$~erg~s$^{-1}$. Most of these
sources lie within the disk of M81, and about one third of sources
lie outside the $D_{25}$ isophote and fourteen ones of them fall
within the IPN box. There are four sources which coincide with
supernova remnants in the disk of M81 and one with the supernova
SN1993J. It should be noted that SGR in the Galaxy are being
intensively studied as persistent sources of soft X-ray radiation,
with luminosities of $10^{34}$--$10^{35}$~erg~s$^{-1}$
\citep{Gogus02, Mereghetti06}. Such a luminosity lies below the
threshold of sensitivity of the Chandra observatory in the survey of
M81.

Thus, all the data presented above definitely indicate that apart
from massive atomic hydrogen clouds, in the space between the M81,
M82, and NGC~3077 galaxies a widespread stellar population exists,
including neutron stars. It is quite probable that the formation of
a larger part of these stars was induced by a close approach of the
galaxies about 300 million years ago.

\begin{figure}
\centering
\includegraphics[width=12cm]{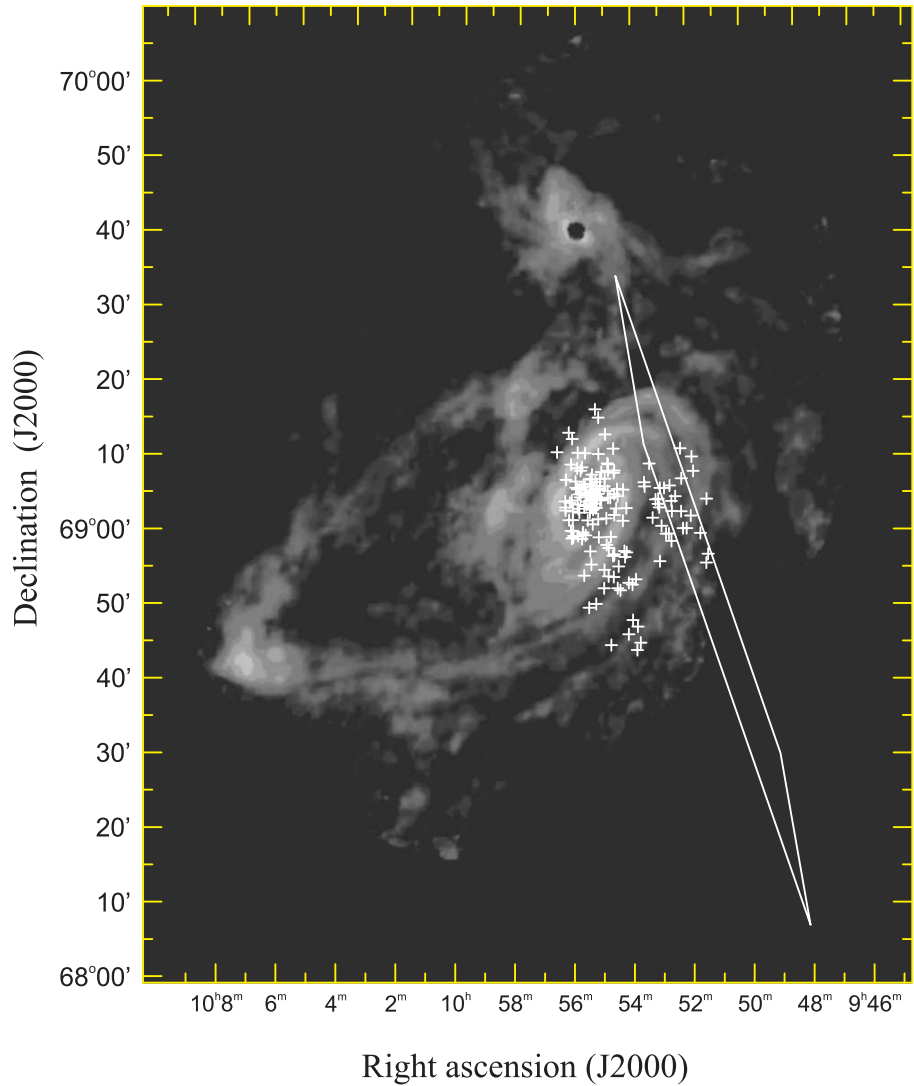}
\caption{The 21~cm HI emission map of the central region of the M81
group of interacting galaxies. M81 at the center; M82 $\sim
35\arcmin$ to the north; and NGC~3077 $\sim 40\arcmin$ to the east
and $\sim 20\arcmin$ to the south. X-ray sources (crosses) observed
by Chandra, and IPN box of GRB~051103 are superimposed. \label{map}}
\end{figure}

Figure~\ref{map} shows the 21-cm map\footnote{See Yun,
www.astro.umass.edu/$\sim$myun/m81hi.html} of the M81/M82/NGC~3077
group, on which the IPN box of GRB~051103 \citep{Golenetskii05} is
overplotted and the positions of the X-ray sources discovered in the
M81 galaxy and its vicinity \citep{Swartz03} are marked. The figure
clearly demonstrates that the IPN box envelops the region that
contains a considerable fraction of the intergalactic HI and
intergalactic stellar population of M81. This eliminates the main
objections to the possible association of the GRB~051103 burst with
the giant flare in the M81 group of galaxies and strongly enhances
the plausibility of this association. If the flare is emitted by a
soft gamma repeater in the M81 group, its energy will come to $\sim7
\times 10^{46}$~erg. This value exceeds the flare energy in
SGR~1806-20 by only a factor of 2--3.

To draw more decisive conclusions, additional studies of IPN box
area in the optical, X-ray, and radio wave bands are necessary.

\acknowledgements This work has been supported by Russian Space
Agency contract and RFBR grant 06-02-16070.

\pagebreak
%

%

\clearpage
\begin{table}[t]
\vspace{6mm} \centering \caption{Error box of GRB~051103
localization} \label{tab}

\vspace{5mm}
\begin{tabular}{|r|c|c|} \hline
 & RA (J2000) & Dec (J2000) \\
\hline
Box center & $9^{\mathrm h} 52\fm5$ & $68\arcdeg 51\farcm6$ \\
Corners 1   & $9^{\mathrm h}56\fm0$ & $69\arcdeg 34\farcm0$ \\
 2   & $9^{\mathrm h} 54\fm2$ & $69\arcdeg 11\farcm3$ \\
 3   & $9^{\mathrm h} 51\fm0$ & $68\arcdeg 30\farcm0$ \\
 4   & $9^{\mathrm h} 50\fm2$ & $68\arcdeg 07\farcm0$ \\
\hline
\end{tabular}
\end{table}

\clearpage
\clearpage
\clearpage
\clearpage
\end{document}